# Linking electronic transport through a spin crossover thin film to the molecular spin state using X-ray absorption spectroscopy operando techniques

*Filip Schleicher*[1], *Michał Studniarek*[1], *Kuppusamy Senthil Kumar*[1], *Etienne Urbain*[1], *Kostantine Katcko*[1], *Jinjie Chen*[2], *Timo Frauhammer*[2], *Marie Hervé*[2], *Ufuk Halisdemir*[1], *Lalit Mohan Kandpal*[1], *Daniel Lacour*[3], *Alberto Riminucci*[4], *Loic Joly*[1], *Fabrice Scheurer*[1], *Benoit Gobaut*[5], *Fadi Choueikani*[5], *Edwige Otero*[5], *Philippe Ohresser*[5], *Jacek Arabski*[1], *Guy Schmerber*[1], *Wulf Wulfhekel*[2,6], *Eric Beaurepaire*[1], *Wolfgang Weber*[1], *Samy Boukari*[1], *Mario Ruben*[1,6], *Martin Bowen*[1,*]

[1]IPCMS UMR 7504 CNRS, Université de Strasbourg, 23 Rue du Loess, BP 43, 67034 Strasbourg Cedex 2, France

[2]Physikalisches Institut, Karlsruhe Institute of Technology, Wolfgang-Gaede-Str. 1, 76131 Karlsruhe, Germany

[3]Institut Jean Lamour UMR 7198 CNRS, Université de Lorraine, BP 70239, 54506 Vandoeuvre les Nancy, France

[4]ISMN-CNR, Via Gobetti 101, 40129 Bologna, Italy

[5]Synchrotron SOLEIL, L'Orme des Merisiers, Saint-Aubin - BP 48, 91192 Gif-sur-Yvette, France

[6]Institute of Nanotechnology, Karlsruhe Institute of Technology (KIT), Hermann-von-Helmholtz-Platz 1, 76344 Eggenstein-Leopoldshafen, Germany

**Corresponding author:**

*bowen@unistra.fr

One promising route toward encoding information is to utilize the two stable electronic states of a spin crossover molecule. However, while this property is clearly manifested in transport across single molecule junctions, evidence linking charge transport across a solid-state device to the molecular film's spin state has thus far remained indirect. To establish this link, we deploy materials-centric and device-centric operando experiments involving X-ray absorption spectroscopy. We find a correlation between the temperature dependencies of the junction resistance and the Fe spin state within the device's $Fe(bpz)_2(phen)$ molecular film. We also factually observe that the Fe molecular site mediates charge transport. Our dual operando studies reveal that transport involves a subset of molecules within an electronically heterogeneous spin crossover film. Our work confers an insight that substantially improves the state-of-the-art regarding spin crossover-based devices, thanks to a methodology that can benefit device studies of other next-generation molecular compounds.

**TOC GRAPHICS**

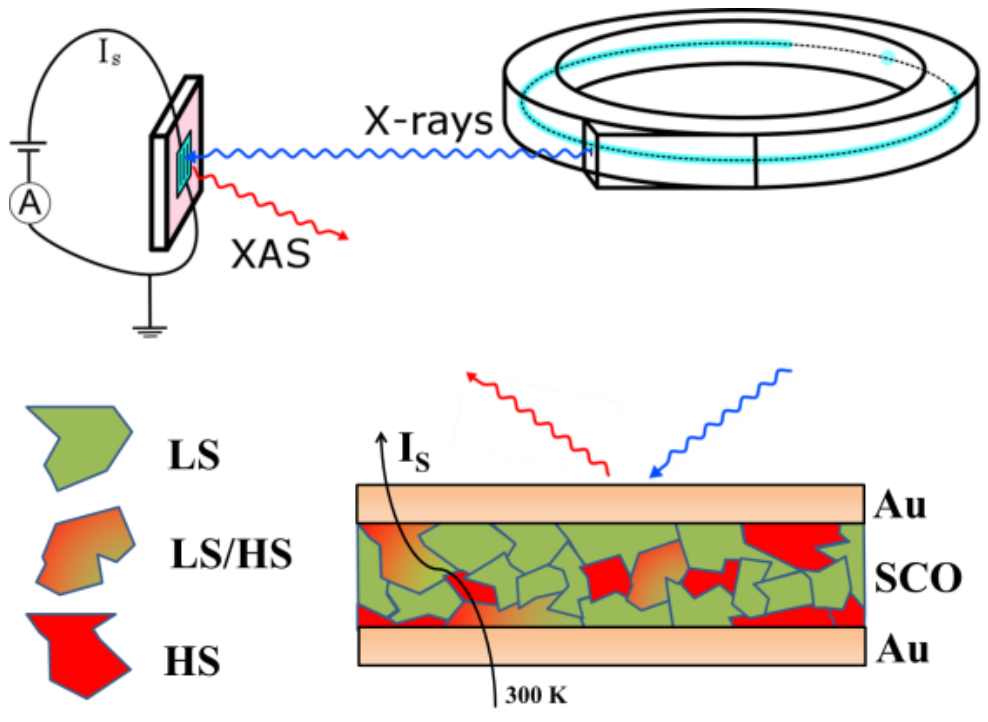

**KEYWORDS** Spin crossover thin film, solid-state device, operando measurements, spintronics, X-ray absorption spectroscopy

A promising route to overcoming the miniaturization limits of traditional storage technologies involves the use of molecules whose conformation, and resulting electronic properties, can be switched between two stable states. One such family of molecular candidates is that of spin crossover (SCO) molecules[1,2], in which the ligand field around a central metal ion may be reversibly altered so as to switch the atom's electronic configuration between low-spin (LS) and high-spin (HS) states. This conformational change occurs via external stimuli (e.g. temperature, light, electric field), which would confer multifunctionality to a SCO-based device[3].

A substantial difficulty toward successfully assembling devices with a bistability inherited from a molecule's SCO property is that the electronically fragile SCO property can become frozen upon forming an interface with the device electrode. In particular, the large interfacial charge transfer that underscores this change can be mitigated by considering surfaces with a reduced density of states and/or reduced chemical activity[4–6]. The SCO transition temperature may also be manipulated through suitable surface engineering[7–11].

Witnessing how the SCO property impacts a device's response has been explicitly achieved through experiments on model junctions assembled using the tip of a scanning tunnelling microscope [2,3]. However, transposing this level of knowledge to solid-state device research, i.e. with industrial prospects, has remained challenging[1,3]. A few reports of device research involving SCO molecules have provided phenomenological modelling of experimental transport data to support a picture of thermal[12–15], electrical field[16] or optical[17] SCO-induced changes to device operation. However, an explicit link between electronic transport and the actual fraction $\rho_{HS}$ of molecules in the HS state within the device, let alone within the actual charge transport path across the SCO film has, to the best of our knowledge, not yet been demonstrated.

Establishing this link is best achieved through an *in operando* technique. Two approaches are possible. In a so-called 'materials-centric' operando approach, the device is placed in a given state and the materials property is read out through the standard materials science method. However, the atoms probed by this materials science technique do not necessarily contribute equally to a device's operation. To improve causality between materials science and device research, a 'device-centric' operando approach can focus this materials characterization onto those atoms that drive the device's operation by examining the materials property *within* device operation (e.g. current flow)[18].

In this paper, we use synchrotron-grade materials-centric and device-centric *operando* measurements[18] combining the element specificity of X-ray absorption spectroscopy (XAS) and electrical transport in order to examine how the thermally activated spin transition impacts transport across a vertical crossbar device based on the generic $Fe(bpz)_2(phen)$ (bpz = dihydrobis(pyrazolyl)borate, phen = 1,10-phenanthroline) family of molecules. Using the materials-centric technique, we find that the temperature dependence of the fraction $\rho_{HS}$ of molecules in the HS state within a Au/[$Fe(H_2B(pz)_2)_2(NH_2$-phen)]/Au trilayer, extracted from XAS measurements, tracks the temperature dependence of junction resistance. Device-centric operando measurements confirm that the Fe central site is involved in transport, and shed light on the subset of SCO molecules that are actually involved in SCO-driven transport across the electronically heterogeneous SCO film.

We chose the $Fe(bpz)_2(phen)$ family of molecules because not only can it be sublimed[19], but it also shows a high degree of functionalization potential, and its members can be synthesized using easily accessible and inexpensive precursors. The molecule studied in this work is [$Fe(H_2B(pz)_2)_2(NH_2$-phen)], with $NH_2$-phen = 5-amino1,10-phenanthroline (see inset to Fig. 1a). Thanks to the addition of the $NH_2$ functional group, the studied compound may easily act as a frame for the realization of functional SCO molecules. Details of the molecular synthesis are described in Supplementary note 1. We show in Fig. 1a the temperature dependence of the magnetization ($\chi_m T$) of a powder reference acquired using superconducting quantum interference device magnetometry (SQUID) upon cooldown and warmup. SQUID magnetometry shows a nearly vanishing $\chi_m T$ at low temperatures that reflects the non-magnetic (S=0) LS state, and a thermal SCO to the HS state at a critical transition temperature of 154 K for the bulk powder without any apparent thermal hysteresis ($\Delta T$) .

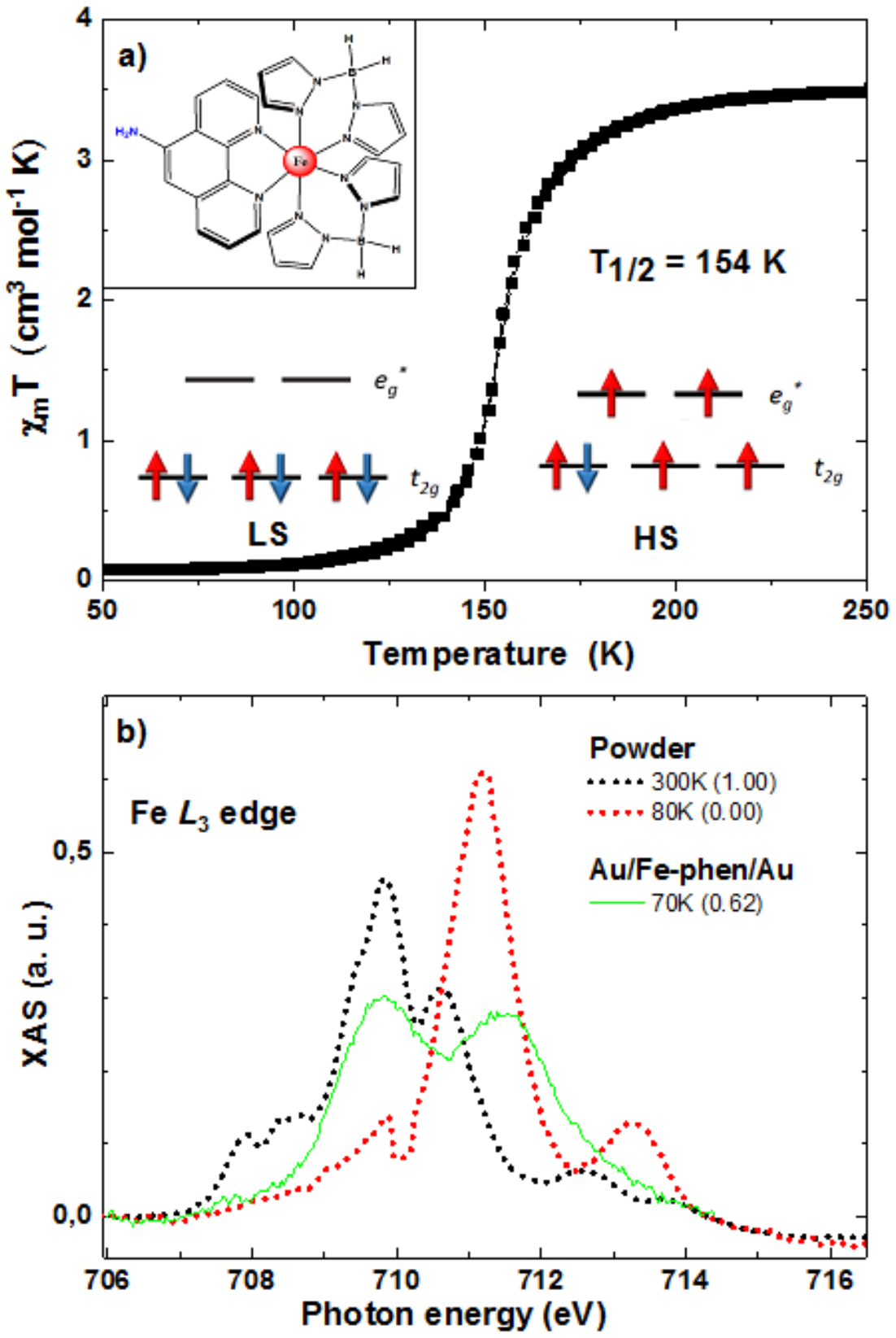

**Figure 1.** (a) χmT vs. T plot of the $[Fe(H_2B(pz)_2)_2(NH_2\text{-phen})]$ powder reference and corresponding electronic configuration for the HS and LS state. (b) X-ray absorption spectra at the Fe $L_3$ edge of the powder reference in the HS- (at 300 K, black; $\rho_{HS} \equiv 1$) and LS-state (at 70 K, red; $\rho_{HS} \equiv 0$) , and of the Au/$[Fe(H_2B(pz)_2)_2(NH_2\text{-phen})]$/Au trilayer at 70 K (green line), for which a value of the trilayer $\rho_{HS} = 0.62$ is extracted by using the peak intensity ratio method (see Supplementary Note 2).

Trilayer device stacks incorporating $[Fe(H_2B(pz)_2)_2(NH_2\text{-phen})]$ molecules were fabricated on thermally oxidized Si/$SiO_x$ substrates. Prior to the deposition, the substrates were cleaned in an oxygen plasma. The entire structure consisted of Au(20nm)/$[Fe(H_2B(pz)_2)_2(NH_2\text{-phen})]$(42nm)/Au(20nm), in which the Au electrodes were evaporated in UHV through a shadow mask, with ex-situ transfer to a current-heated crucible molecular evaporator. Molecules were sublimed at a pressure P = 4.3E-8 mbar without the mask and covered the entire sample. The intersection between the bottom and top Au electrodes with a respective width of 300 μm and 50 μm thus defined the junctions. This junction size represents a compromise between a reasonably high device resistance (typically ~6 MΩ at low bias) and continuity of electrodes evaporated through the mask atop the SCO film.

The device was measured on the DEIMOS beamline[20] of Synchrotron SOLEIL using the $V^2TI$ electrical insert[21]. Junctions were bonded in 2-point mode, and measured using a Keithley 2636 Sourcemeter in voltage source mode. The X-ray beam impinged normal to the sample surface. The dataset of Fig. 2 was acquired in a materials-centric[18] operando mode: we witnessed the impact of varying the sample temperature on the SCO property by monitoring XAS through total fluorescence yield (TFY) at the Fe $L_3$ edge, and by measuring the junction resistance independently. Since molecules cover the entire sample surface, and since our 600x800 $\mu m^2$ beam cannot fit within the 300x50 $\mu m^2$ junction size, in order to probe only molecules that are sandwiched between Au electrodes, our materials-centric data was acquired atop a mm-sized trilayer stack next to the junction.

We present in Fig.1b Fe $L_3$ edge XAS reference powder spectra in the HS (300 K, black dots; $\rho_{HS} \equiv 1$) and LS (80 K, red dots; $\rho_{HS} \equiv 0$) states. These spectra exhibit peaks at 709.8 eV and 711.3 eV corresponding to Fe $t_{2g}4e_g2$ ($S$ = 2) and $t_{2g}6e_g0$ ($S$ = 0) configurations for HS and LS molecules, respectively[22]. We also show a XAS spectrum acquired at 70 K on the Au/[$Fe(H_2B(pz)_2)_2(NH_2$-phen)]/Au trilayer (green line) after subtracting a baseline. By fitting this spectrum using a peak intensity ratio method based on the HS and LS reference spectra of the powder sample, we extracted a proportion $\rho_{HS}$ of molecules in the HS state of 0.61 at 70 K for molecules sandwiched between Au electrodes in our devices[11] (see Supplementary Note 2). By repeating this fitting procedure on XAS spectra acquired on the reference SCO junction upon varying the sample temperature, we can then extract the temperature dependence of the trilayer $\rho_{HS}$ for molecules sandwiched between Au electrodes. We plot it in Fig. 2a (green squares), alongside the temperature dependence of the reference $\rho_{HS}$ for both the powder sample (filled dots) and a 230nm-thick film deposited on gold (open black dots).

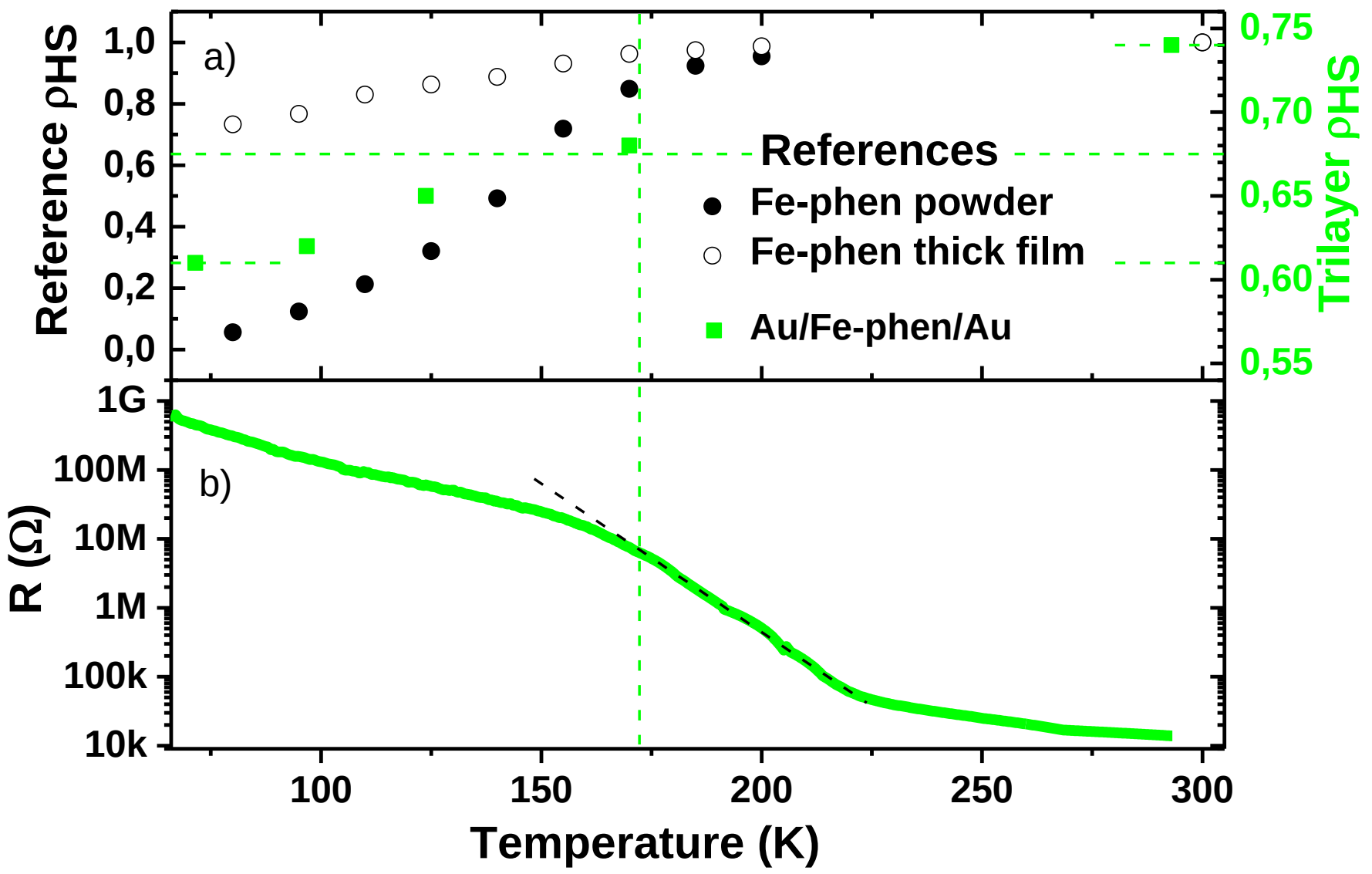

**Figure 2.** Temperature dependencies of (a) the HS proportion $\rho_{HS}$ in a Au/[Fe($H_2B(pz)_2)_2(NH_2$-phen)]/Au trilayer(green), a [Fe($H_2B(pz)_2)_2(NH_2$-phen)] reference powder (filled black) and a 230 nm-thick [Fe($H_2B(pz)_2)_2(NH_2$-phen)] film deposited onto Au (open black), and (b) of the Au/[Fe($H_2B(pz)_2)_2(NH_2$-phen)]/Au device resistance R. To ensure a measurable current at all temperatures, a voltage of 800 mV was applied. The green dashed lines peg the ~170 K temperature at which the trilayer $\rho_{HS}$ has decreased by half of its total amplitude due to the thermally activated SCO. This coincides with a deviation in $R(T)$ from the exponential dependence of $R(T$>170 K) (intersection of vertical dashed green line with black dashed line).

We find that, at room temperature, $\rho_{HS}$ = 0.74, i.e only 74% of molecules sandwiched between gold electrodes are in the HS state, while at 70 K $\rho_{HS}$ drops to 0.61. This suggests that only 13% of the molecules actually switch, implying that the remaining molecules are blocked in the HS and LS states (61% and 26%, respectively). This can arise from the aforementioned electronic coupling to the gold electrodes[23–25] ($\rho_{HS}$ only reaches 0.75 at 80 K in the thick reference film deposited on gold, while it is ~0 at 80K for the powder) or from modified intermolecular interactions in the molecular film[26]. These aspects can also account for the different transition temperatures observed in the powder, the thick film deposited on gold and the thin film sandwiched between Au electrodes[11]. Finally, the gradual character of the transition suggests reduced cooperativity of the molecules with respect to the powder sample[8,26,27].

Turning now to Fig. 2b, the temperature dependence of the SCO device resistance R generally increases as temperature is decreased (note the resistance log scale used). Starting from 300K, this increase is nearly exponential, which is typical of thermally assisted hopping transport across a SCO organic layer[13]. Starting at roughly 220K, this exponential increase becomes steeper, but is arrested around 175 K. It is also at this temperature that the trilayer $\rho_{HS}$ has decreased by half of its total amplitude (see dashed lines in panels (a) and (b) of Fig. 2). For temperatures lower than around 175 K, another exponential dependence appears to drive the temperature dependence of *R*. The data also suggest that, setting aside thermally assisted hopping, the LS state *R* is lower than the HS state *R*, in line with other studies[13,17,28].

This comparison between the temperature dependencies of $\rho_{HS}$ and of device resistance in terms of two HS- and LS-dominated transport regimes thus suggests that transport across the junction indeed involves the SCO property. However, since our XAS-based determination of $\rho_{HS}$ remains 'materials-centric', it does not explicitly probe the spin state of the molecules that form the actual charge transport path across the SCO film. To further support this claim of SCO-driven operation, we examine in a 'device-centric' operando approach[18] if the Fe site of our [$Fe(H_2B(pz)_2)_2(NH_2$-phen)] SCO molecule is involved in transport. To do so, we positioned the X-ray beam on the junction and swept the photon energy across the Fe $L$ edge while simultaneously measuring TFY (see Fig.3a) and raw current flow across the junction for ±10 mV (see Fig.3b) at $T$ = 300 K.

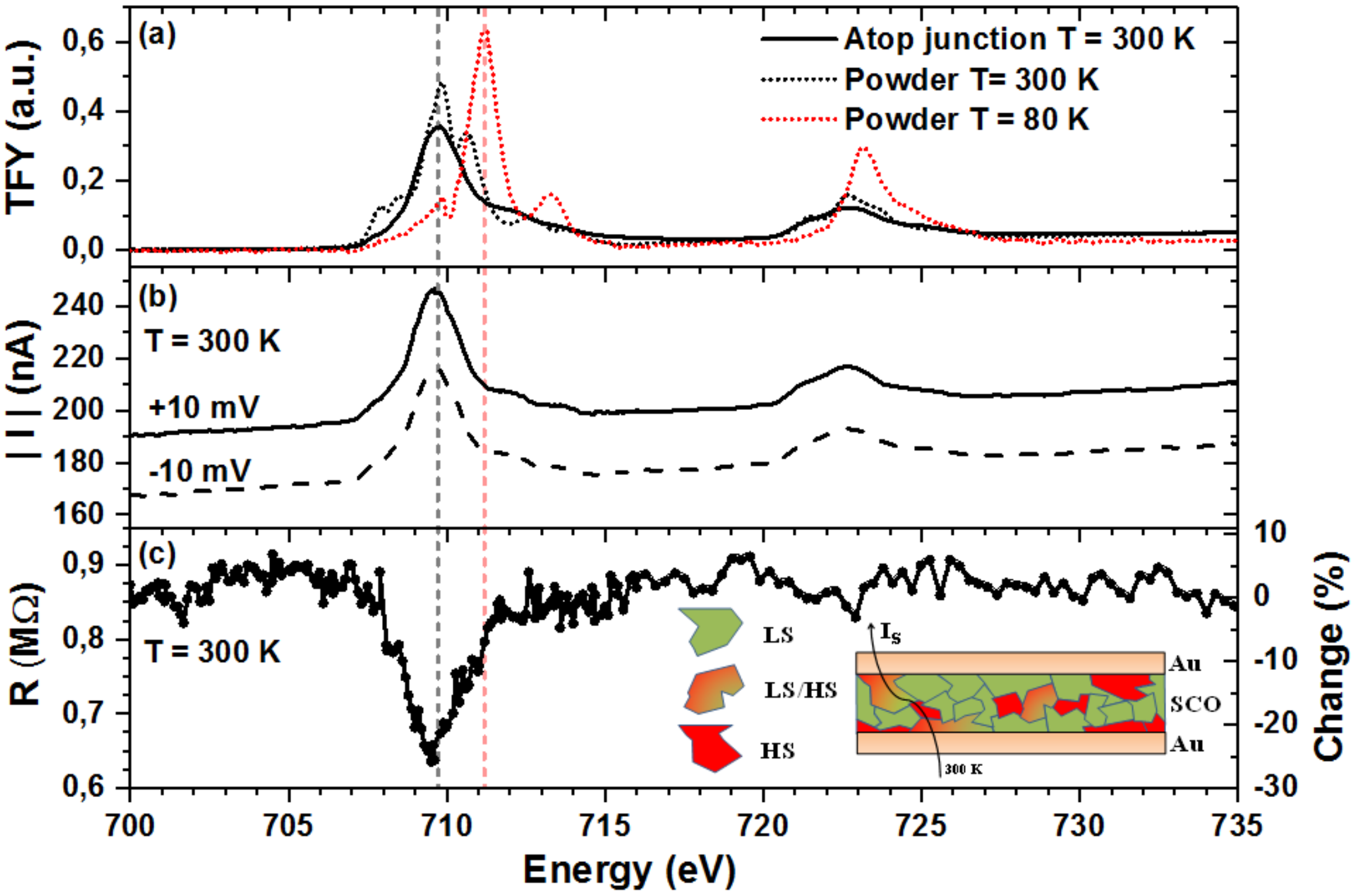

**Figure 3.** Impact at $T$ = 300 K of sweeping photon energy across the Fe $L_3$ edge on a) X-ray absorption spectra measured on the 300x50 $\mu m^2$ Au/[$Fe(H_2B(pz)_2)_2(NH_2$-phen)]/Au junction in TFY mode with a 800x600 $\mu m^2$ X-ray beam, and on b) the current flowing across the device for both signs of 10mV applied bias, i.e in a device-centric operando approach[11]. The HS state witnessed in the TFY spectrum of panel (a) corroborates data for the thick SCO reference film of

Fig. 2a. Panel (a) also shows reference powder spectra in the HS and LS states. |I| is plotted in panel (b) to facilitate the inspection of the ±10 mV data. c) Impact of Fe $L_3$ edge absorption on the device resistance after removing the photocurrent. The resistance at 300 K drops by 25 % upon reaching the Fe $L_3$ edge, while the spectrum mimicks the XAS spectrum of the HS state. The inset to panel c) schematizes how the current (black arrow) flows across HS sites that are either frozen (red zones) or switchable (red/green zones) at 300 K and not across LS sites (green zones).

Both the TFY and raw junction current spectra clearly reveal the impact of X-ray absorption by Fe sites. Given the respective 800x600 $\mu m^2$ and 300x50 $\mu m^2$ sizes of the X-ray beam and the junction, the TFY spectrum of Fig. 3a mostly probes molecules outside the junction, and the HS state witnessed in the TFY spectrum corroborates data for the thick reference SCO film of Fig. 2a. This also means that the raw current spectra of panel (b) naturally include a photocurrent resulting from the absorption of X-rays by Fe sites belonging to molecules that are predominantly outside the junction. Here, the sample is grounded through our IV measurement unit. To remove this photocurrent, so as to concentrate on device transport, we assume a bias-symmetric operation at low bias and subtract the two ±10 mV raw current spectra[18]. The resulting photon energy dependence of device resistance is plotted in Fig. 3c. We find a 25% drop in device resistance upon reaching the Fe $L_3$ edge maximum at 709.8eV, which corresponds to the HS state[29].

Our 'device-centric' operando data thus indicate that the transport path at $T$ = 300 K involves only molecules in the HS state. At low temperature, the raw device resistance was unmeasurable at the low 10mV bias required[18] by the 'device-centric' operando technique. We were thus unable to characterize the transport path at low temperature. Nevertheless, as mentioned before, the LS R is reported[2] to be lower than the HS R, and $\rho_{HS}$ = 0.74 at $T$ = 300 K within the trilayer according to the 'materials-centric' operando data. Combining information accrued from our 'materials-centric' and 'device-centric' operando experiments, we draw the following picture of the heterogeneous electronic and transport properties across the Au/[$Fe(H_2B(pz)_2)_2(NH_2$-phen)]/Au junction. 1) A 0.26 proportion of molecules is frozen in the LS state at T = 300 K and does not contribute to device operation at T = 300K. 2) Of the 0.74 proportion of Fe sites in the HS state at T = 300K, only a 0.13 proportion manifest the SCO property, while a 0.61 proportion of molecules remain frozen in the HS state. As mentioned before, this freezing of the spin state into LS and HS configurations can arise from electronic coupling to the gold electrodes, from modified intermolecular interactions, and also potentially from structural degradation. 3) Transport at 300K proceeds across molecules in the HS state (see inset to Fig. 3c)). 4) While molecules that are in the frozen HS state could be contributing here, we witness how only a 0.13 thermally driven reduction to $\rho_{HS}$, which probes all Fe sites, can nevertheless heavily impact the temperature dependence of the junction resistance. The transport path therefore contains molecules with an active SCO property (see red/green zones within nanoscale transport path schematized in Fig 3c).

In conclusion, using X-ray absorption spectroscopy (XAS), we performed a material-centric and device-centric operando study on junctions integrating a $Fe(bpz)_2$(phen) derivative spin crossover thin film. Using materials-centric XAS scans, we extracted the temperature dependence $\rho_{HS}(T)$ of the high-to-low spin ratio for a [$Fe(H_2B(pz)_2)_2(NH_2$-phen)] film sandwiched between Au electrodes. We find that $\rho_{HS}(T)$ can explain features of the junction resistance's temperature dependence directly in terms of the high-to-low spin ratio of molecules between the Au electrodes. Using the device-centric XAS operando technique, we confirmed the involvement of Fe sites within the charge transport path across the vertical device's SCO layer. By comparing the materials-centric and device-centric data, we qualitatively establish a picture of SCO-driven transport across a subset of the molecules forming the electronically heterogeneous SCO film within the device's trilayer structure, with an insight that substantially improves the state-of-the-art regarding SCO-based solid-state device knowledge. Our work should stimulate further studies, within a challenging context, to enhance the impact of the SCO property on the operation/performance of solid-state devices. A particularly promising prospect remains to successfully combine the SCO property with the high spin polarization at the ferromagnet/molecule interface[30], e.g. by tailoring the charge transfer thanks to a suitable noble metal spacer[31]. Finally, our work outlines a powerful methodology to examine how the operation of solid-state devices is impacted by the properties of next-generation molecular compounds incorporated therein.

ASSOCIATED CONTENT

**Supporting Information.**

Synthesis of $[Fe(H_2B(pz)_2)_2(NH_2\text{-phen})]$, HS/LS ratio determination (single PDF)

AUTHOR INFORMATION

**Notes**

The authors declare no competing financial interests.

ACKNOWLEDGMENTS

We acknowledge funding from the Franco-German university, from the Baden-Württemberg Stiftung in the framework of the Kompetenznetz für Funktionale Nanostrukturen (KFN) and from the Agence Nationale de la Recherche (ANR-11-LABX-0058 NIE; "SpinMarvel" ANR-09-JCJC-0137 ; "Spinapse" ANR-14-CE26-0009-01), from the Institut Carnot MICA's 'Spinterface' grant and from the International Center for Frontier Research in Chemistry and from the Deutsche Forschungsgemeinschaft (DFG grant Wu349/13-1). We acknowledge Synchrotron SOLEIL for provision of synchrotron radiation facilities and we would like to thank the SOLEIL staff for assistance in using beamline DEIMOS (proposal number 20160249)

REFERENCES

(1) Molnár, G.; Rat, S.; Salmon, L.; Nicolazzi, W.; Bousseksou, A. Spin Crossover Nanomaterials: From Fundamental Concepts to Devices. *Adv. Mater.* **2017**, 17003862.

(2) Kumar, K. S.; Ruben, M. Emerging Trends in Spin Crossover (SCO) Based Functional Materials and Devices. *Coord. Chem. Rev.* **2017**, *346*, 176–205.

(3) Lefter, C.; Davesne, V.; Salmon, L.; Molnár, G.; Demont, P.; Rotaru, A.; Bousseksou, A. Charge Transport and Electrical Properties of Spin Crossover Materials: Towards Nanoelectronic and Spintronic Devices. *Magnetochemistry* **2016**, *2* (1), 18.

(4) Bernien, M.; Wiedemann, D.; Hermanns, C. F.; Krüger, A.; Rolf, D.; Kroener, W.; Müller, P.; Grohmann, A.; Kuch, W. Spin Crossover in a Vacuum-Deposited Submonolayer of a Molecular Iron(II) Complex. *J. Phys. Chem. Lett.* **2012**, *3* (23), 3431–3434.

(5) Miyamachi, T.; Gruber, M.; Davesne, V.; Bowen, M.; Boukari, S.; Joly, L.; Scheurer, F.; Rogez, G.; Yamada, T. K.; Ohresser, P.; et al. Robust Spin Crossover and Memristance across a Single Molecule. *Nat. Commun.* **2012**, *3*, 938.

(6) Bairagi, K.; Iasco, O.; Bellec, A.; Kartsev, A.; Li, D.; Lagoute, J.; Chacon, C.; Girard, Y.; Rousset, S.; Miserque, F.; et al. Molecular-Scale Dynamics of Light-Induced Spin Cross-over in a Two-Dimensional Layer. *Nat. Commun.* **2016**, *7*, 12212.

(7) Zhang, X.; Costa, P. S.; Hooper, J.; Miller, D. P.; N'Diaye, A. T.; Beniwal, S.; Jiang, X.; Yin, Y.; Rosa, P.; Routaboul, L.; et al. Locking and Unlocking the Molecular Spin Crossover Transition. *Adv. Mater.* **2017**, *29* (39), 1702257.

(8) Davesne, V.; Gruber, M.; Studniarek, M.; Doh, W. H.; Zafeiratos, S.; Joly, L.; Sirotti, F.; Silly, M. G.; Gaspar, A. B.; Real, J. A.; et al. Hysteresis and Change of Transition Temperature in Thin Films of $Fe\{[Me_2Pyrz]_3BH\}_2$, a New Sublimable Spin-Crossover Molecule. *J. Chem. Phys.* **2015**, *142* (19), 194702.

(9) Mikolasek, M.; Félix, G.; Nicolazzi, W.; Molnár, G.; Salmon, L.; Bousseksou, A. Finite Size Effects in Molecular Spin Crossover Materials. *New J. Chem.* **2014**, *38* (5), 1834.

(10) Félix, G.; Nicolazzi, W.; Salmon, L.; Molnár, G.; Perrier, M.; Maurin, G.; Larionova, J.; Long, J.; Guari, Y.; Bousseksou, A. Enhanced Cooperative Interactions at the Nanoscale in Spin-Crossover Materials with a First-Order Phase Transition. *Phys. Rev. Lett.* **2013**, *110* (23).

(11) Kumar, K. S.; Studniarek, M.; Heinrich, B.; Arabski, J.; Schmerber, G.; Bowen, M.; Boukari, S.; Beaurepaire, E.; Dreiser, J.; Ruben, M. Engineering On-Surface Spin Crossover: Spin-State Switching in a Self-Assembled Film of Vacuum-Sublimable Functional Molecule. *Adv. Mater.* **2018**, 1705416.

(12) Shi, S.; Schmerber, G.; Arabski, J.; Beaufrand, J.-B.; Kim, D. J.; Boukari, S.; Bowen, M.; Kemp, N. T.; Viart, N.; Rogez, G.; et al. Study of Molecular Spin-Crossover Complex Fe(Phen)2(NCS)2 Thin Films. *Appl. Phys. Lett.* **2009**, *95* (4), 043303.

(13) Devid, E. J.; Martinho, P. N.; Kamalakar, M. V.; Šalitroš, I.; Prendergast, Ú.; Dayen, J.-F.; Meded, V.; Lemma, T.; González-Prieto, R.; Evers, F.; et al. Spin Transition in Arrays of Gold Nanoparticles and Spin Crossover Molecules. *ACS Nano* **2015**, *9* (4), 4496–4507.

(14) Takahashi, K.; Cui, H.-B.; Okano, Y.; Kobayashi, H.; Einaga, Y.; Sato, O. Electrical Conductivity Modulation Coupled to a High-Spin−Low-Spin Conversion in the Molecular System $[Fe^{III}(Qsal)_2][Ni(Dmit)_2]_3 \cdot CH_3CN \cdot H_2O$. *Inorg. Chem.* **2006**, *45* (15), 5739–5741.

(15) Prins, F.; Monrabal-Capilla, M.; Osorio, E. A.; Coronado, E.; van der Zant, H. S. J. Room-Temperature Electrical Addressing of a Bistable Spin-Crossover Molecular System. *Adv. Mater.* **2011**, *23* (13), 1545–1549.

(16) Meded, V.; Bagrets, A.; Fink, K.; Chandrasekar, R.; Ruben, M.; Evers, F.; Bernand-Mantel, A.; Seldenthuis, J. S.; Beukman, A.; van der Zant, H. S. J. Electrical Control over the Fe(II) Spin Crossover in a Single Molecule: Theory and Experiment. *Phys. Rev. B* **2011**, *83* (24).

(17) Lefter, C.; Rat, S.; Costa, J. S.; Manrique-Juárez, M. D.; Quintero, C. M.; Salmon, L.; Séguy, I.; Leichle, T.; Nicu, L.; Demont, P.; et al. Current Switching Coupled to Molecular Spin-States in Large-Area Junctions. *Adv. Mater.* **2016**, *28* (34), 7508–7514.

(18) Studniarek, M.; Halisdemir, U.; Schleicher, F.; Taudul, B.; Urbain, E.; Boukari, S.; Hervé, M.; Lambert, C.-H.; Hamadeh, A.; Petit-Watelot, S.; et al. Probing a Device's Active Atoms. *Adv. Mater.* **2017**, *29* (19), 1606578.

(19) Naggert, H.; Bannwarth, A.; Chemnitz, S.; von Hofe, T.; Quandt, E.; Tuczek, F. First Observation of Light-Induced Spin Change in Vacuum Deposited Thin Films of Iron Spin Crossover Complexes. *Dalton Trans.* **2011**, *40* (24), 6364.

(20) Ohresser, P.; Otero, E.; Choueikani, F.; Chen, K.; Stanescu, S.; Deschamps, F.; Moreno, T.; Polack, F.; Lagarde, B.; Daguerre, J.-P.; et al. DEIMOS: A Beamline Dedicated to Dichroism Measurements in the 350–2500 EV Energy Range. *Rev. Sci. Instrum.* **2014**, *85* (1), 013106.

(21) Joly, L.; Muller, B.; Sternitzky, E.; Faullumel, J.-G.; Boulard, A.; Otero, E.; Choueikani, F.; Kappler, J.-P.; Studniarek, M.; Bowen, M.; et al. Versatile Variable Temperature Insert at the DEIMOS Beamline for *in Situ* Electrical Transport Measurements. *J. Synchrotron Radiat.* **2016**, *23* (3), 652–657.

(22) Bernien, M.; Naggert, H.; Arruda, L. M.; Kipgen, L.; Nickel, F.; Miguel, J.; Hermanns, C. F.; Krüger, A.; Krüger, D.; Schierle, E.; et al. Highly Efficient Thermal and Light-Induced Spin-State Switching of an Fe(II) Complex in Direct Contact with a Solid Surface. *ACS Nano* **2015**, *9* (9), 8960–8966.

(23) Gruber, M.; Miyamachi, T.; Davesne, V.; Bowen, M.; Boukari, S.; Wulfhekel, W.; Alouani, M.; Beaurepaire, E. Spin Crossover in Fe(Phen) $_2$ (NCS) $_2$ Complexes on Metallic Surfaces. *J. Chem. Phys.* **2017**, *146* (9), 092312.

(24) Gopakumar, T. G.; Bernien, M.; Naggert, H.; Matino, F.; Hermanns, C. F.; Bannwarth, A.; Mühlenberend, S.; Krüger, A.; Krüger, D.; Nickel, F.; et al. Spin-Crossover Complex on Au(111): Structural and Electronic Differences Between Mono- and Multilayers. *Chem. - Eur. J.* **2013**, *19* (46), 15702–15709.

(25) Shepherd, H. J.; Molnár, G.; Nicolazzi, W.; Salmon, L.; Bousseksou, A. Spin Crossover at the Nanometre Scale. *Eur. J. Inorg. Chem.* **2013**, *2013* (5–6), 653–661.

(26) Naggert, H.; Rudnik, J.; Kipgen, L.; Bernien, M.; Nickel, F.; Arruda, L. M.; Kuch, W.; Näther, C.; Tuczek, F. Vacuum-Evaporable Spin-Crossover Complexes: Physicochemical Properties in the Crystalline Bulk and in Thin Films Deposited from the Gas Phase. *J. Mater. Chem. C* **2015**, *3* (30), 7870–7877.

(27) Gütlich, P.; Hauser, A.; Spiering, H. Thermal and Optical Switching of Iron(II) Complexes. *Angew Chem. Int Ed Engl.* **1994**, *33*, 2024–2054.

(28) Studniarek, M. Interface and Multifunctional Device Spintronics: Studies with Synchrotron Radiation, Strasbourg, 2016.

(29) The absence of a visible resistance drop at the Fe $L_2$ edge in Fig. 3c likely reflects the lower $L_2$ amplitude and the signal-to-noise ratio of our resistance data.

(30) Djeghloul, F.; Gruber, M.; Urbain, E.; Xenioti, D.; Joly, L.; Boukari, S.; Arabski, J.; Bulou, H.; Scheurer, F.; Bertran, F.; et al. High Spin Polarization at Ferromagnetic Metal–Organic Interfaces: A Generic Property. *J. Phys. Chem. Lett.* **2016**, *7* (13), 2310–2315.

(31) Gruber, M.; Ibrahim, F.; Boukari, S.; Joly, L.; Da Costa, V.; Studniarek, M.; Peter, M.; Isshiki, H.; Jabbar, H.; Davesne, V.; et al. Spin-Dependent Hybridization between Molecule and Metal at Room Temperature through Interlayer Exchange Coupling. *Nano Lett.* **2015**, *15* (12), 7921–7926.